\documentclass{myPoS}
\usepackage{graphicx}
\usepackage{subfigure}
\usepackage{cite,mcite}

\usepackage{xspace}

\def\Ord{\lower .7ex\hbox{$\;\stackrel{\textstyle <}{\sim}\;$}}
\def\OOrd{\lower .7ex\hbox{$\;\stackrel{\textstyle >}{\sim}\;$}}

\def\ttbar{$t\overline{t}$\xspace}%

\title{Cross-sections and branching ratios for charged Higgs searches}

\ShortTitle{Cross-sections and branching ratios for charged Higgs searches}

\author{\speaker{Andr\'e Sopczak}\\
        Lancaster / Uppsala University\\
        E-mail: \email{andre.sopczak@cern.ch}}

\abstract{For the preparation of the experimental search for charged  
Higgs bosons at the LHC detailed studies have been made to  
determine the expected charged Higgs boson production  
cross-sections and decay branching ratios at $\sqrt{s}=14$~TeV. In the mass  
regime below the $t$-quark mass the expected production cross-sections  
are discussed using PYTHIA and FeynHiggs program packages 
based on the decay $t\to H^+b$. 
For higher masses Next-to-Leading-Order (NLO) calculations have  
been used, and particular attention has been given to the intermediate-mass 
region. The decay branching ratios have been studied with the  
program packages FeynHiggs and HDecay. Higher-order  
corrections ($\Delta_b$ corrections) in the MSSM are consistently taken  
into account. Two benchmark scenarios are considered, one of them the  
`mhmax'. }

\FullConference{Contribution to the Workshop on \\
	        Prospects for Charged Higgs Discovery at Colliders\\
		16-19 September 2008, Uppsala, Sweden,\\
	        to be published by http://pos.sissa.it}

\begin{document}


\section{Introduction}

Charged Higgs bosons are naturally predicted in non-minimal Higgs scenarios, 
for instance in Two Higgs Doublet Models (THDMs), and specifically in the
Minimal Supersymmetric Standard Model (MSSM)~\cite{guide}.

At present, a lower bound on the charged Higgs boson mass of about 80~GeV exists 
from LEP~\cite{leplimit,as2006},
remarkably close to the previously simulated sensitivity~\cite{lep_hpm_mass}.
With initial Tevatron data, upper limits were placed on 
BR$(t \to H^+b)$ for different charged Higgs decay scenarios~\cite{cdf,phil}.
Starting from 2009/2010, the LHC at CERN will enable the discovery or the ruling out of the 
existence of such a particle over a large portion 
of both the THDM and MSSM parameter space (masses up to around 400~GeV).
The exact value of the reach depends on the value of $\tan\beta$ 
(reviews~\cite{Roy:2004az,Roy:2004mm,Roy:2005yu} and a recent study~\cite{Mohn:2007fd}).
The expected sensitivities for the LHC have been reported at this workshop~\cite{martin,ritva}.
 
This comparative study focuses on the latest developments in the production cross-section
and branching ratio determination, and is structured as follows. 
First the parameter points in the MSSM are defined in section~2. 
Then in section~3 the determination of the branching ratio BR($t\to H^+b$) 
is addressed and results from PYTHIA and FeynHiggs calculations are discussed.
This branching ratio is used to determine the $H^+$ production cross-section
in the low-mass region (section~4). Section~5 addresses the 
intermediate-mass region around $m_{H^+}=170$~GeV. In the high-mass
region the process $gb\to H^+$ is dominant and the calculations are
described in section~6. The $H^+$ branching ratios are discussed in
section~7. Systematic uncertainties are the focus in section~8. Section~9
describes the data-base structure for storing the cross-section and branching
ratio values.

\section{MSSM scenarios}
Two scenarios in the MSSM are considered. They are described by the following 
parameters.
Scenario A:
$m_{t} = 175$~GeV,
$M_{\rm SUSY} = 500$~GeV,
$A_t = 1000$~GeV,
$\mu = 200$~GeV,
$M_2 = 1000$~GeV,
$M_3 = 1000$~GeV.
Scenario B (``mhmax'')~\cite{mhmax}:
$m_{t} = 175$~GeV,
$M_{\rm SUSY} = 1000$~GeV,
$X_t = 2000$~GeV, where $A_t = X_t + \mu/\tan\beta$,
$\mu = 200$~GeV,
$M_2 = 200$~GeV,
$M_3 = 800$~GeV.

The $\Delta_b$ corrections are calculated in FeynHiggs v2.6.2~\cite{feynhiggs} for these two cases
in the $H^+$ couplings.
For $\tan\beta =50$ they are $\Delta_b=0.45$ for scenario A, and $\Delta_b=0.21$ for scenario B.
The $\Delta_b$ corrections modify the $b$-quark mass $m_b^{\rm corrected}=m_b/(1+\Delta_b)$~\cite{Carena:1999py}.

\section{$\mathbf{t\to H^+b}$ branching ratios}

\begin{figure}[h!]
\begin{center}
\includegraphics[width=0.49\textwidth]{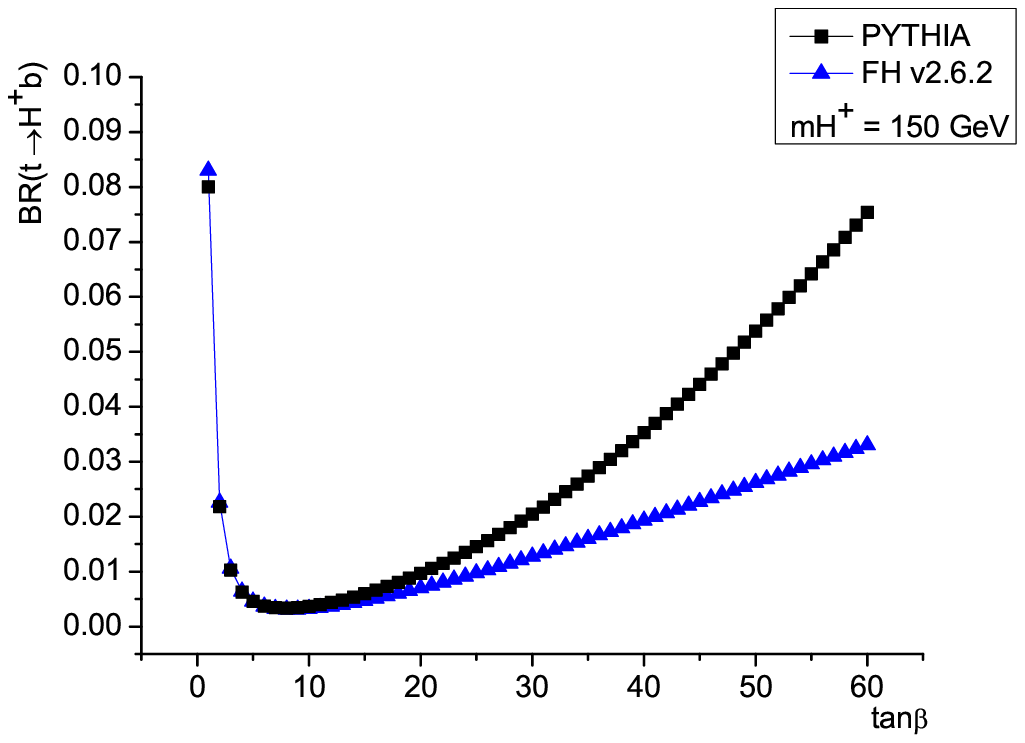} \hfill
\includegraphics[width=0.49\textwidth]{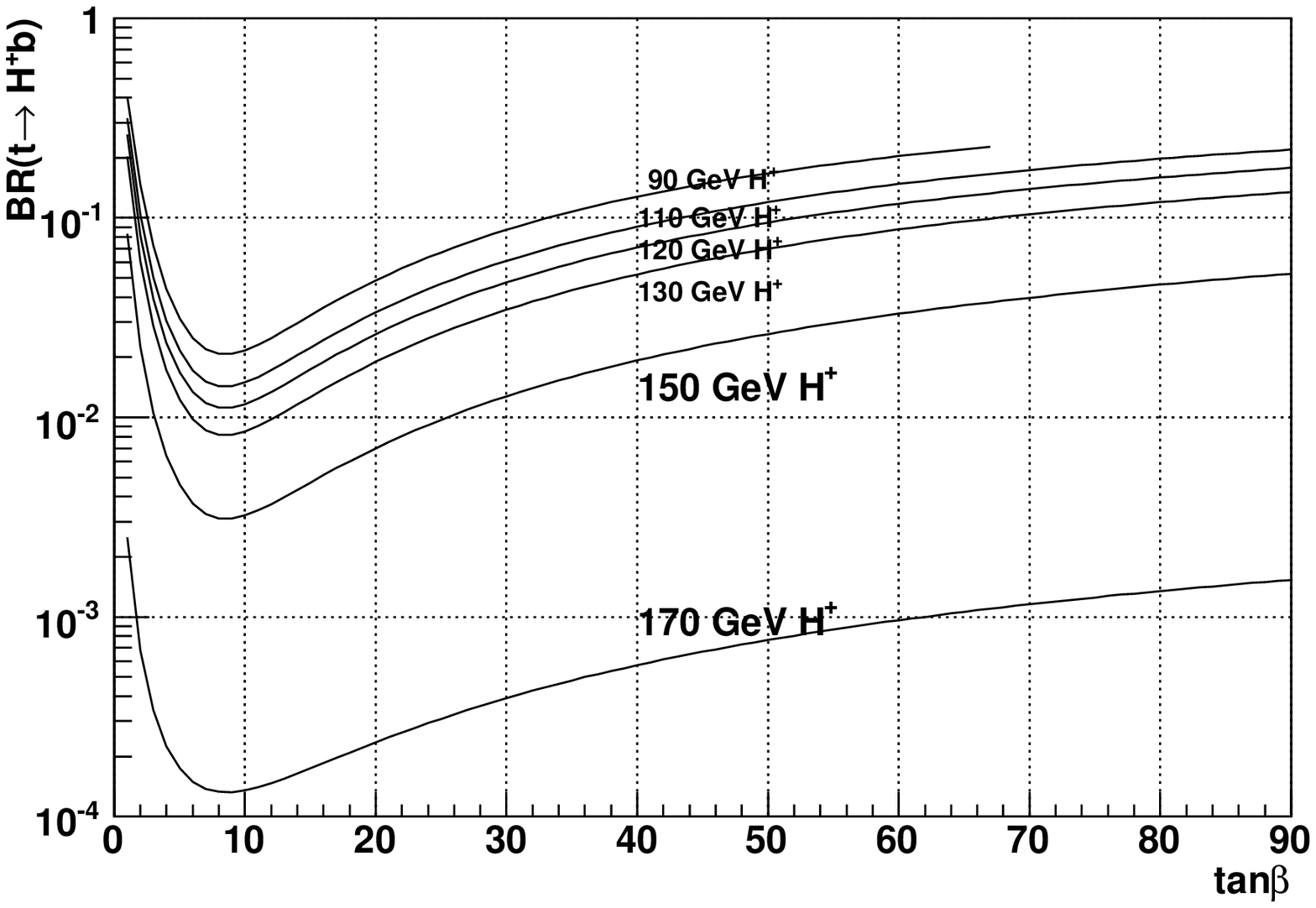}
\vspace*{-0.2cm}
\caption{\small Left: Expected branching ratio BR($t\to H^+b$) for MSSM scenario A.
Right: 
Expected branching ratio BR($t\to H^+b$) for MSSM scenario A
as described in the text, calculated with FeynHiggs version~2.6.2.
For very large $\tan\beta$ values and low charged Higgs masses (90 GeV), 
the model enters a non-perturbative regime and loop calculations are no 
longer valid, as indicated by the ending of the 90~GeV curve.
\label{fig:br_tbh150}}
\vspace*{-0.7cm}
\end{center}
\end{figure}

The ${\rm BR}(t\to H^+b)$ values have been determined with PYTHIA v6.325~\cite{pythia}
and FeynHiggs v2.6.2~\cite{feynhiggs}\footnote{In this study the branching ratios 
were also produced with FeynHiggs v2.6, however, then discarded as the differences 
in version 2.6 and 2.6.2 were only attributed to a programming correction 
(`bug fix') in the latter version.}.
In FeynHiggs the formula from Ref.~\cite{top_to_hb} is implemented, and furthermore it includes
the $\Delta_b$ corrections depending on the MSSM parameters~\cite{Carena:1999py}.
The computations have been performed for MSSM scenarios A and B.
An example is shown in Fig.~\ref{fig:br_tbh150} for scenarios A as a function of $\tan\beta$.
The FeynHiggs calculations include $\Delta_b$ corrections, while the PYTHIA calculation 
does not. The $\Delta_b$ corrections are positive and grow with $\tan\beta$ which explains the observed
increase in BR($t\to H^+b)$ difference with increasing $\tan\beta$ (Fig.~\ref{fig:br_tbh150}).

The first step in the determination of the production cross-section is the 
calculation of the branching ratio BR($t\to H^+b$) in the low mass region 90 to 170~GeV.
These branching ratios are also shown in Fig.~\ref{fig:br_tbh150} for scenario A, 
calculated with FeynHiggs.

\section{Low-mass cross-section}

The charged Higgs boson production cross-section has been evaluated with different program packages
for low-mass and high-mass regions. If the charged Higgs boson mass $m_{H^+}$ satisfies 
$m_{H^+} < m_{t} - m_{b}$, where $m_{b}$ is the $b$-quark mass,
$H^+$ particles could be produced in the on-shell process $ t \rightarrow bH^+$,
the latter being in turn produced from $gg$ fusion  and $q\bar q$ annihilation.
Contributions of gluon fusion to the production of charged Higgs at hadron colliders
were pointed out previously~\cite{lorenzo}.
This approximation has customarily been used in event generators for $m_{H^+} \Ord m_{t}$.

Charged Higgs production is denoted by $q\overline{q}$, $ gg \rightarrow t\overline{t} \rightarrow tbH^+$ 
if due to (anti-) top decays  and by $q\overline{q}$, $ gg \rightarrow tbH^+$ if further production diagrams 
are included~\cite{Alwall:2003tc}. Owing to the large $t$-quark decay width ($ \Gamma_{t} \simeq 1.5$~GeV) and 
to  the additional diagrams which do not proceed via direct $ t\overline{t}$ 
production~\cite{Borzumati:1999th,Miller:1999bm,Moretti:1999bw}, charged Higgs bosons could 
also be produced at and beyond the mass threshold ($m_t-m_b$) in the $tbH^+$ process. 
The importance of these effects in the so-called `threshold' or 
`transition'  region (intermediate-mass region $m_{H^+}\approx m_t$) was emphasized in 
various Les Houches proceedings~\cite{Cavalli:2002vs,Assamagan:2004mu}
as well as in Refs.~\cite{Alwall:2003tc,Guchait:2001pi,Moretti:2002ht,Assamagan:2004gv}, 
and the calculations of Refs.~\cite{Borzumati:1999th,Miller:1999bm} 
(based on the appropriate $q\overline{q},gg\to tb H^+$ description) 
are now implemented in 
HERWIG\,\cite{herwig,Corcella:2000bw,Corcella:2002jc,Moretti:2002eu}\,and 
PYTHIA~\cite{pythia,Alwall:2003tc}. A comparison between
the two generators was carried out in Ref.~\cite{Alwall:2003tc}.
In addition,
in the mass region near the $t$-quark mass, a matching of the calculations for the
$ q\overline{q},~gg \rightarrow tbH^+$ and 
$ gb \rightarrow tH^+$ processes is required already in leading order~\cite{Alwall:2004xw}.
In the kinematic region where $t\to H^+b$ is possible, this process could dominate the charged 
Higgs production.

The cross-sections in the low-mass region for the charged Higgs boson masses
90, 110, 120, 130 and 150~GeV have been calculated from 
the higher-order improved \ttbar cross-section ($\sigma_{t\bar t}=833$~pb~\cite{ttxsection}) and
the ${\rm BR}(t\to H^+b)$ determined from FeynHiggs (version 2.6.2)~\cite{feynhiggs}:
$\sigma_{tbH^+}=2\cdot\sigma_{t\bar t}{\rm BR}(t\to H^+b)[1-{\rm BR}(t\to H^+b)].$
Results are shown in Fig.~\ref{fig:factor} (right plot).

\section{Intermediate-mass cross-section}
 
Charged Higgs bosons with a mass of 170 GeV would predominantly be produced by the 
$gb \to tH^+$ process. 
The intermediate-mass region has been studied in NLO~\cite{berger}.
The $t$-quark mass has been fixed to 175~GeV.
For this scenario with a 5 GeV mass difference between charged Higgs 
and $t$-quark masses, the additional cross-section from the $t\to H^+b$ process amounts to 
an increase of about 20 to 30\%. We have taken into account this increase in the 
derivation of the production cross-section by adding both cross-sections.
The cross-section increase depends strongly on the mass difference between charged Higgs 
and $t$-quark masses, and also on the treatment of the running $b$-quark mass.
Results are shown in Fig.~\ref{fig:factor} (right plot, 170~GeV curve).

\section{High-mass cross-section}

At hadron colliders, the main contribution to charged Higgs boson production is 
through the twin processes $gg \to tbH^+$ and $gb \to tH^+$ for $m_{H^+} > m_t$. 
These are called twin processes since they correspond to two different approximations 
describing the same basic process. For charged Higgs boson masses above the $t$-quark 
mass, the $2\to 2$ process is dominant, due to the resummation of potentially large 
logarithms in the $b$-quark parton density~\cite{tilman}.
In the high-mass region the Next-to-Leading Order (NLO) production
cross-section calculation is applied~\cite{berger,tilman}.
In this case, the parton shower produces an 
outgoing $b$-quark of relative small transverse momentum. 
In the region of phase space where the outgoing $b$-quark has large 
transverse momentum, the parton shower does not give a good description of the process, 
and the full $2\to 3$ description is needed. However, these two descriptions overlap 
for small transverse momenta of the $b$-quark, necessitating a matching procedure to remove 
resulting double counting~\cite{Alwall:2004xw}.
For charged Higgs boson masses below the $t$-quark mass, the $2\to 3$ process dominates 
since it incorporates on-shell $t$-quark pair production with subsequent decay into a 
charged Higgs boson.
Matchig~\cite{matching2} is a new leading-order event generator based on the work presented 
in Ref.~\cite{Alwall:2004xw} which matches 
the two processes by producing negative weight events from an identified double-counting term.
In the high-mass region the NLO program package~\cite{berger} also avoids double counting.

\begin{figure}[h!]
\begin{center}
\vspace*{-0.3cm}
\includegraphics[width= 0.49\textwidth]{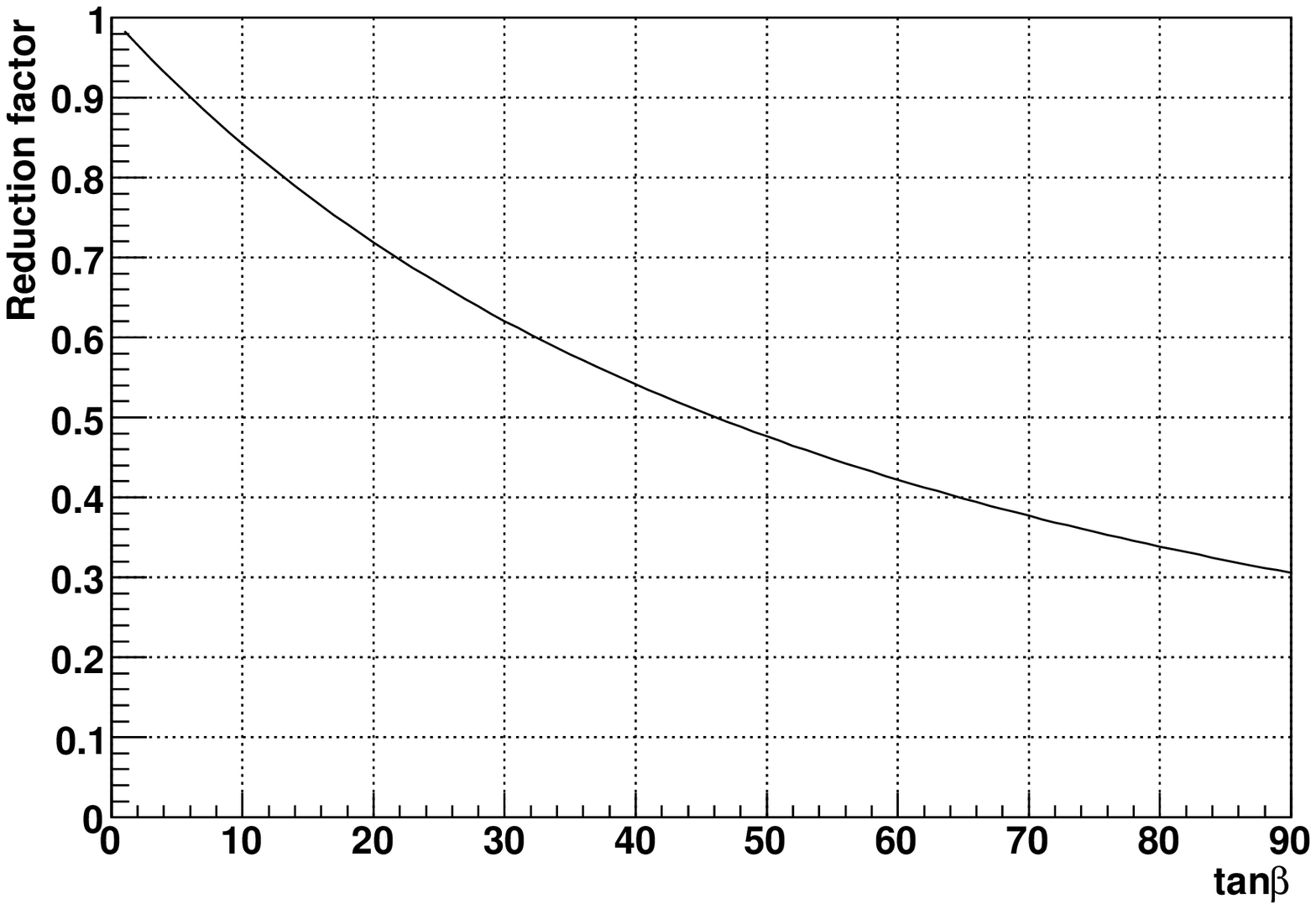}  \hfill
\includegraphics[width= 0.49\textwidth]{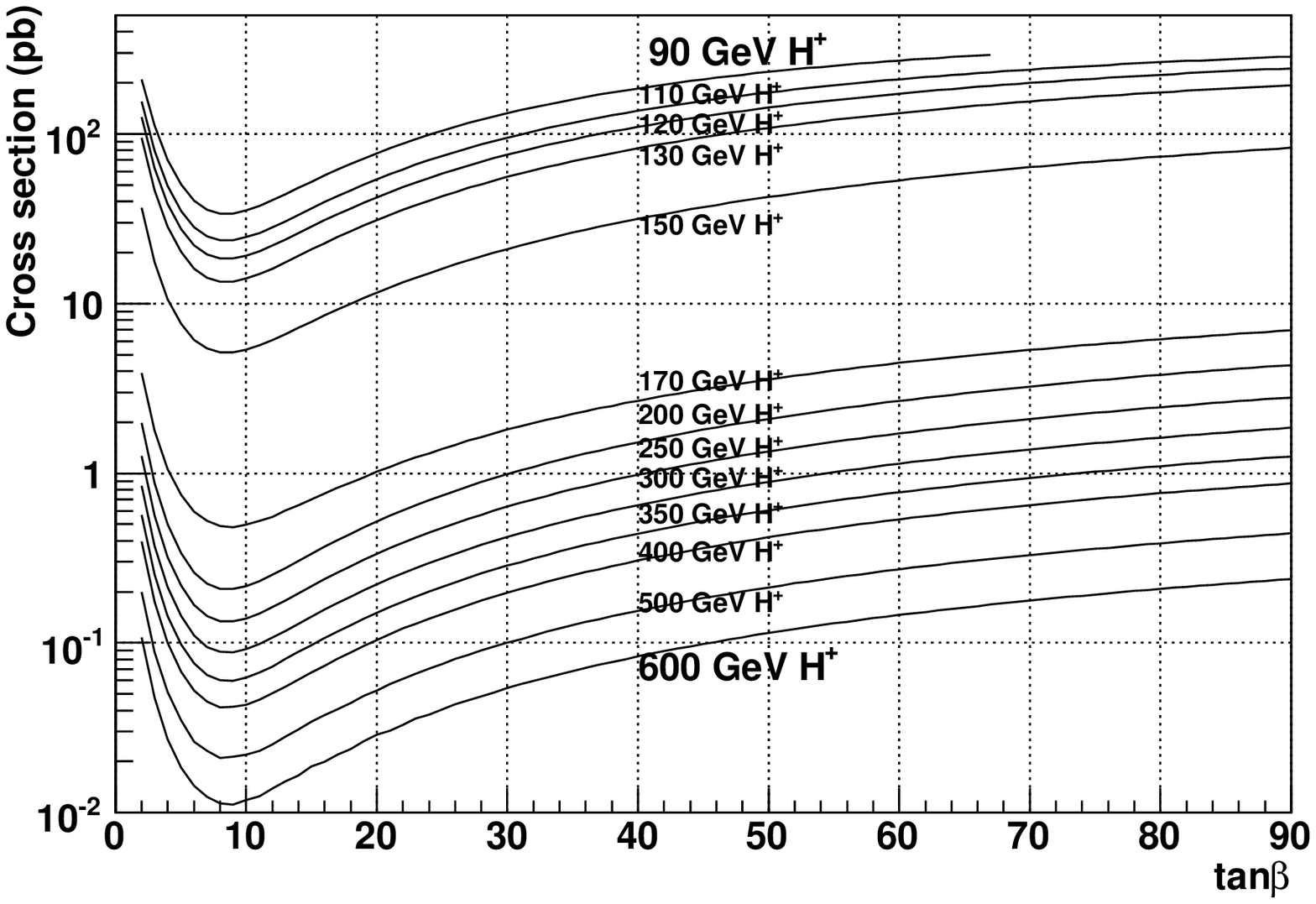}  \hfill
\vspace*{-0.2cm}
\caption{\small Left: Reduction factors $f$ which have been applied to the NLO charged Higgs boson 
                production cross-section in the MSSM for scenario A.
         The reduction factors result from $\Delta_b$ corrections and have been calculated with
         FeynHiggs v2.6.2~\protect\cite{feynhiggs}.
         Right: Expected charged Higgs boson production cross-section in the MSSM for scenario A.
         For charged Higgs boson masses of 170~GeV and above the reduction factors $f$ have been applied 
         to the NLO cross-section calculations~\protect\cite{tilman}. This allows a consistent treatment
         of the $\Delta_b$ corrections for the production cross-sections and branching ratios over the 
         whole mass range.
\label{fig:factor}}
\vspace*{-0.8cm}
\end{center}
\end{figure}

For the charged Higgs boson masses 200, 250, 350, 400, 500 and 600~GeV, the ${\rm BR}(t\to H^+b)$
is kinematically suppressed and the cross-sections have been determined in NLO from the 
$gb\to tH^+$ process alone~\cite{tilman}.
We have explicitly not calculated the dependence of the cross-sections on the MSSM 
parameters with the NLO program package~\cite{berger,tilman}\footnote{Recently, 
the charged Higgs production process at NLO has been implemented in the program package PROSPINO~2.1 
including $\Delta_b$ corrections.}.
These higher-order corrections can be large,
depending on the MSSM scenario. They depend primarily on $\tan\beta$. In order to determine 
these corrections, first the $\Delta_b$ corrections are calculated with the FeynHiggs
package~\cite{feynhiggs}. Then a reduction factor $f=1/(1+\Delta_b)^2$ 
is calculated for the production cross-section.
The previously determined NLO cross-sections are multiplied by this reduction factor.
The reduction factors for scenario A are shown in Fig.~\ref{fig:factor} as a function of $\tan\beta$.
The figure shows also the cross-sections after the application of the reduction 
factors\footnote{However, for any sub-dominant decay channel of a heavy charged Higgs boson, such as 
                 $H^+\to\tau^+\nu$, $\Delta_b$ corrections cancel to a large extent~\cite{eriksson}.}.

\section{$\mathbf{H^+}$ branching ratios}

This section describes the $H^+$ branching ratio studies, in particular
focusing on the off-mass-shell effects and the $\Delta_b$ corrections.
The ${\rm BR}(H^+\to ...)$ values have been determined with the FeynHiggs v2.5, v2.6.2 and HDecay 3.2 program packages.
FeynHiggs v2.5 does not include off-mass-shell effects while FeynHiggs v2.6.2 does.
The difference between FeynHiggs v2.6.2 and HDecay 3.2\footnote{Since version 3.3 $\Delta_b$ corrections are included 
in HDecay for the charge Higgs sector.} is that FeynHiggs includes the $\Delta_b$ corrections.

While for $m_{H^+}\Ord m_{t}$ charged Higgs bosons decay 
predominantly into a $\tau$-lepton and a neutrino, 
or into a $cs$-quark pair;
for $m_{H^+}\OOrd m_{t}$ both $H^+\to \tau\nu_\tau$ and 
$H^+\to tb$ are important decay channels.
In the experimental search, the latter is much harder to disentangle from background than the former.  
The associated $t$-quark decays predominantly into a $W$ boson
and a $b$-quark.

The branching ratios have been determined with the FeynHiggs~\cite{feynhiggs} and 
HDecay~\cite{hdecay} program packages. The detailed comparison showed very good agreement
between FeynHiggs and HDecay 
calculations for the branching ratios in the low-mass region
($m_{H^+} <m_t$), including the virtual effects which lead to $H^+\to tb$ contribution in this 
mass region\footnote{These virtual effects were only included in FeynHiggs version 2.6, release 
July 2007. In previous versions of FeynHiggs the $tb$ branching fraction due to virtual 
effects was zero.}. In the high-mass region,
vertex corrections ($\Delta_b$ terms) which are included in FeynHiggs lead to a significant
variation of the branching ratio with $\tan\beta$ while the branching ratios calculated with 
HDecay are largely independent of $\tan\beta$ in the high-mass region. 
The branching ratio comparison is given in Table~\ref{tab:br_comparison}.

\begin{table}[htb]
\begin{center}
\small
\begin{tabular}{|r|rrrrrr|}
\hline
 Decay mode &  BR($H^+\to ...$) $\tau\nu$ & $cs$ &         $tb$ &  $\tau\nu$ &         $cs$ &         $tb$ \\
$m_{H^+}$ (GeV) &        170 &        170 &        170 &        400 &        400 &        400 \\
\hline
 $\tan\beta =3$, FH & 0.901/0.766 & 0.009/0.008 &   0/0.1485 & 0.004/0.004 & 0.000/0.000 & 0.978/0.978 \\
 3, HD &      0.745 &      0.008 &      0.133 &      0.004 &      0.000 & 0.984 \\
 10, FH & 0.990/0.988 & 0.006/0.006 &    0/0.002 & 0.146/0.146 & 0.001/0.001 & 0.845/0.845 \\
 10, HD &      0.974 &      0.006 &      0.001 &      0.112 &          0 &       0.88 \\
 60, FH & 0.991/0.991 & 0.006/0.006 &    0/0.000 & 0.336/0.336 & 0.002/0.001 & 0.660/0.662 \\
 60, HD &      0.976 &      0.006 &          0 &      0.143 &          0 &      0.854 \\
\hline
\end{tabular}  
\caption{\small Branching ratios BR$(H^+\to \tau\nu,~cs,~tb)$ for FeynHiggs (FH), left: v2.5, right v2.6.2,
and HDecay (HD) v3.2. Version 2.6.2 includes off-mass-shell effects, while version v2.5 does not.
It has been explicitly checked that the difference between FeynHiggs v2.6.2 and HDecay are due to
the $\Delta_b$ corrections~\protect\cite{heinemeyer_priv}.
\label{tab:br_comparison}
}
\end{center}
\vspace*{-0.3cm}
\end{table}

\begin{figure}[h!]
\begin{center}
\vspace*{-0.2cm}
\includegraphics[width=0.49\textwidth]{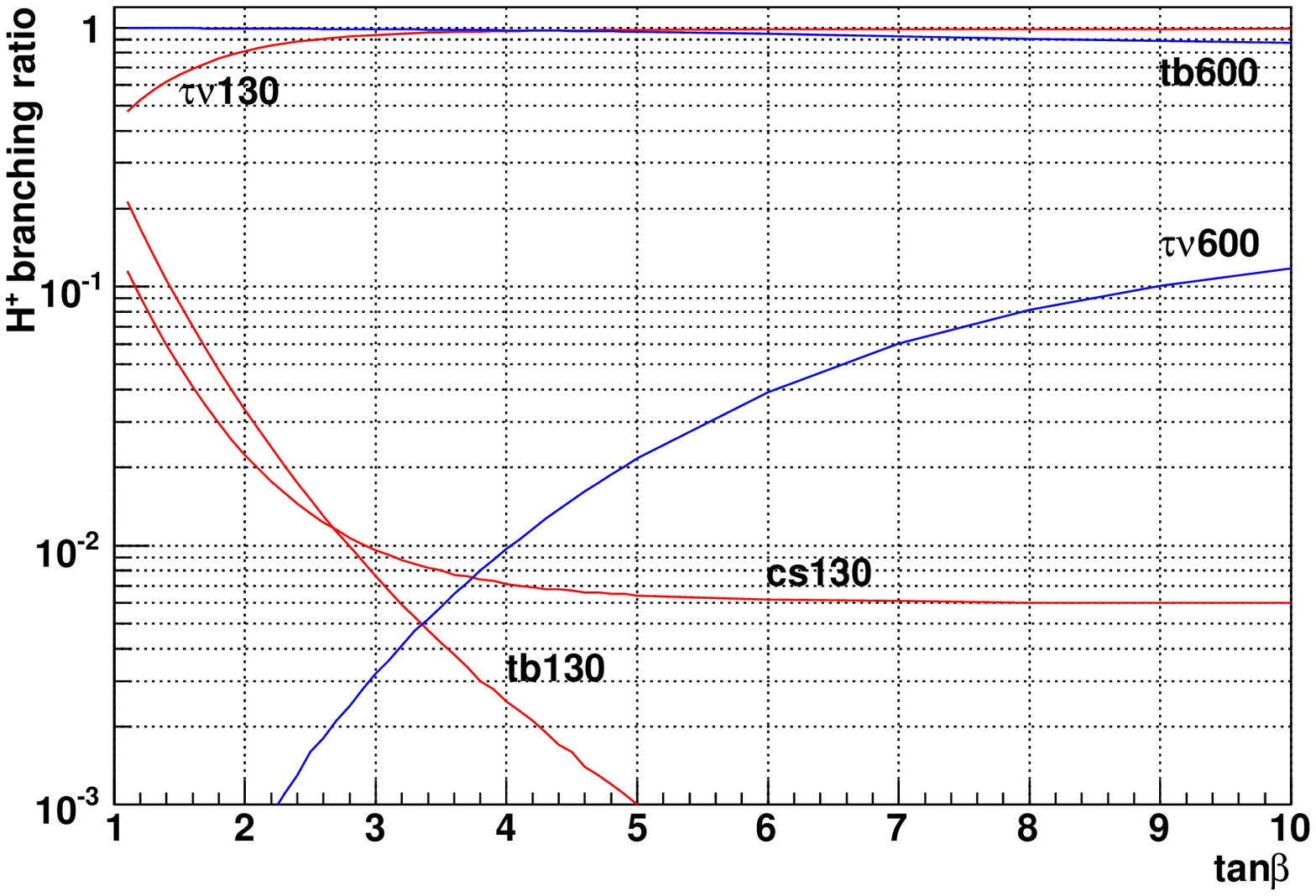}
\includegraphics[width=0.49\textwidth]{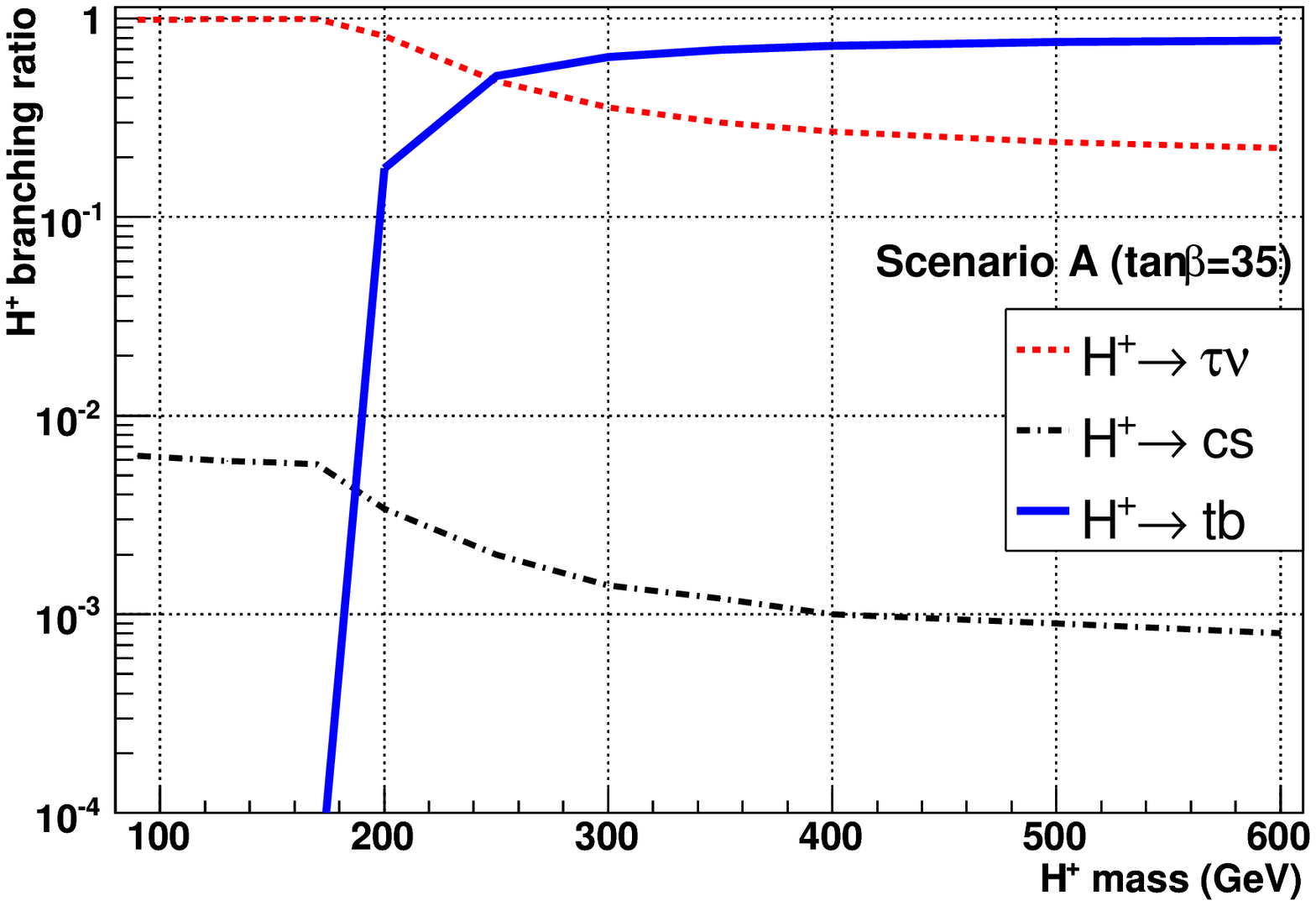}
\vspace*{-0.3cm}
\caption{\small Left: Expected charged Higgs boson branching ratios in the MSSM for scenario A
         for an example of a light
         and a heavy charged Higgs boson~\protect\cite{feynhiggs}. 
Right: Expected charged Higgs boson branching ratios in the MSSM for scenario A
         as a function of the charged Higgs boson mass~\protect\cite{feynhiggs}. 
\label{fig:br}}
\vspace*{-0.5cm}
\end{center}
\end{figure}

Examples of branching ratios are shown in Fig.~\ref{fig:br} 
in the low-mass region for a 130~GeV and in the high-mass region for a 600~GeV charged 
Higgs boson~\cite{feynhiggs}.
The dependence of the branching ratio as a function of the mass is also shown 
in Fig.~\ref{fig:br}.
When kinematically allowed the decay 
$H^+\to\chi^0\chi^+$ 
could become large, as calculated
with FeynHiggs v2.6.2. This decay mode is not addressed in the current discovery analyses
and this branching fraction is not discussed further.

\vspace*{-2mm}
\section{Systematic Uncertainties}
\vspace*{-2mm}

The systematic uncertainties are discussed for the charged Higgs branching ratio
BR($t\to H^+b$) and production cross-section in the low-mass region for $m_{H^+}<m_t$.
The charged Higgs boson branching ratios BR($H^+\to \tau\nu,cs,tb$) have 
been determined with the same program package and similar 
systematic uncertainties apply.
The production cross-sections in the high-mass region have been determined with
a NLO program package~\cite{tilman}, and the systematic uncertainties are discussed separately.

The values for BR($t\to H^+b$) and BR($H^+\to \tau\nu,cs,tb$) have been calculated 
with FeynHiggs v2.6.2 which includes vertex corrections in the framework of the MSSM ($\Delta_b$ terms).
Systematic uncertainties from higher-order loop corrections to the $tbH^+$ vertex
and due to the running of the $c$ and $s$-quark masses are expected~\cite{heinemeyer_priv}.
Upper limits on the resulting systematic uncertainties are conservatively estimated to be:\\
$\Delta {\Gamma}(t\to H^+b)/{\Gamma}<10\%$,
$\Delta {\Gamma}(H^+\to \tau\nu)/{\Gamma}<5\%$,
$\Delta {\Gamma}(H^+\to cs,tb)/{\Gamma}<10\%$.

Systematic uncertainties on the charged Higgs boson production cross-section in the
high-mass region can occur primarily from two sources. First from the renormalization scale
and factorization scale dependence, and second from the fact that supersymmetry loop 
corrections are not (yet) included in the calculations~\cite{tilman03,zhu}.
The one-loop contributions largely improve the theoretical uncertainty of the leading 
order (LO) cross-section. The remaining uncertainty can be estimated from the 
scale dependence. 
The variation of the $gb\to tH^+$ production
cross-section for $0.1<\mu/\mu_{\rm central}<10$ was considered, 
where $\mu_{\rm central}=(m_t+m_{H^+})/5$.
The resulting systematic uncertainty on the production cross-section is below 20\%. 

Corrections from supersymmetric particles in the MSSM are not included in the NLO production
cross-section calculations~\cite{tilman03,zhu}. 
In this study, these supersymmetry loop corrections ($\Delta_b$ corrections) have
been taken into account independent of the NLO calculations using the FeynHiggs package 
and thus no additional uncertainty beyond the 20\% is 
assigned\footnote{Without the $\Delta_b$ corrections, these uncertainties from 
supersymmetric corrections were in addition.
They alter the relation between the bottom mass and the bottom 
Yukawa coupling. These $\Delta m_b$ corrections are the leading supersymmetric one-loop 
corrections with respect to powers of $\tan\beta$. 
(The naming convention for $\Delta m_b$ has changed to $\Delta_b$. 
They refer to the same corrections.)
As pointed out in Ref.~\cite{tilman03,zhu} their 
effect on the total cross-section in a simple mSUGRA model is estimated to stay 
below $\pm 5\%$ for $\tan\beta=30$ and below $\pm 20\%$ for $\tan\beta=50$.
Figure~\ref{fig:factor} of this study shows that the effect can be larger than 50\%
for $\tan\beta>50$ in the MSSM.
In addition to the $\Delta m_b$ corrections higher-order supersymmetry 
QCD corrections have been calculated recently~\cite{spira2008}.}.

\vspace*{-1mm}
\section{Database}
\vspace*{-1mm}

The branching ratio BR$(t\to H^+b)$, the charged Higgs boson production cross-section 
in the low-mass, intermediate-mass and high-mass regions, the $\Delta_b$ values and 
corresponding cross-section reduction factors, and the $H^+$ branching ratios 
have been determined for various charged Higgs boson masses between 90 and 600~GeV
and for $\tan\beta$ values between 1 and 90. 
The calculated values have been stored in a database in root format for scenarios A and B.

\vspace*{-1mm}
\section{Conclusions}
\vspace*{-1mm}

Comparative studies for the expected charged Higgs boson production cross-sections and branching
rations have been performed for searches in the initial LHC data.
The production cross-sections in the low-mass region 
and all charged Higgs branching ratios have been calculated with FeynHiggs 
v2.6.2.
The production cross-sections in the high-mass region have been determined with a
dedicated NLO program package and the dependence on MSSM parameters 
has been added using the FeynHiggs package.

\vspace*{-1mm}
\section*{Acknowledgements}
\vspace*{-1mm}

I would like to thank the organizers of the workshop for their hospitality, in particular, 
Johan Rathsman and Tord Ekel\"of. Also I would like to thank 
Johan Alwall and Oscar St\aa l for fruitful discussion on the charged Higgs production processes,
Tilman Plehn for discussions regarding the NLO cross-section program package,
Michal Spira for discussions on the HDecay program package, 
and Sven Heinemeyer for discussions on the FeynHiggs program package.
I also thank the ATLAS csc HG-10 team for discussions, in particular,
Thies Ehrich,
Martin Flechl, 
Eilam Gross,
Bjarte Mohn and
Remi Zaidan.

\end{document}